\documentclass[aps,pra,twocolumn,superscriptaddress]{revtex4}
\usepackage{amsfonts}
\usepackage{amsmath}
\usepackage{mathrsfs}
\usepackage{amssymb}
\usepackage{graphicx}
\usepackage{dcolumn}
\usepackage{bm}

\begin{document}

\title{Ground-state properties of dipolar Bose-Einstein condensates with
spin-orbit coupling and quantum fluctuations}
\author{Xianghua Su}
\affiliation{Key Laboratory for Microstructural Material Physics of Hebei Province,
School of Science, Yanshan University, Qinhuangdao 066004, China}
\author{Wenting Dai}
\affiliation{Key Laboratory for Microstructural Material Physics of Hebei Province,
School of Science, Yanshan University, Qinhuangdao 066004, China}
\author{Tianyu Li}
\affiliation{Key Laboratory for Microstructural Material Physics of Hebei Province,
School of Science, Yanshan University, Qinhuangdao 066004, China}
\author{Jiyuan Wang}
\affiliation{Key Laboratory for Microstructural Material Physics of Hebei Province,
School of Science, Yanshan University, Qinhuangdao 066004, China}
\author{Linghua Wen}
\email{linghuawen@ysu.edu.cn}
\affiliation{Key Laboratory for Microstructural Material Physics of Hebei Province,
School of Science, Yanshan University, Qinhuangdao 066004, China}

\begin{abstract}
We study the ground-state properties of dipolar spin-1/2 Bose-Einstein
condensates with quantum fluctuations and Rashba spin-orbit coupling (SOC).
The combined effects of dipole-dipole interaction (DDI), SOC, and
Lee-Huang-Yang (LHY) correction induced by quantum fluctuations on the
ground-state structures and spin textures of the system are analyzed and
discussed. For the nonrotating case and fixed nonlinear interspecies contact interaction strengths, our results
show that structural phase transitions can be achieved by adjusting the
strengths of the DDI and LHY correction. In the absence of SOC, a
ground-state phase diagram is given with respect to the DDI strength and the
LHY correction strength. We find that the system exhibits rich quantum
phases including square droplet lattice phase, annular phase, loop-island
structure, stripe-droplet coexistence phase, toroidal stripe phase, and
Thomas-Fermi (TF) phase. For the rotating case, the increase of DDI strength
can lead to a quantum phase transition from superfluid phase to supersolid
phase. In the presence of SOC, the quantum droplets display obvious
stretching and hidden vortex-antivortex clusters are formed in each
component. In particular, weak or moderate SOC favors the formation of
droplets while for strong SOC the ground state of the system develops into a
stripe phase with hidden vortex-antivortex clusters. Furthermore, the system
sustains exotic spin textures and topological excitations, such as
composite skyrmion-antiskyrmion-meron-antimeron cluster, meron-antimeron
string cluster, antimeron-meron-antimeron chain cluster, and peculiar skyrmion-antiskyrmion-meron-antimeron necklace with a meron-antimeron necklace
embedded inside and a central spin Neel domain wall.
\end{abstract}

\maketitle


\section{Introduction}

Dipolar Bose-Einstein condensates (BECs) possess dipole-dipole interaction
(DDI) besides the usual $s$-wave contact interaction. The DDI is long-range
and anisotropic, which has important influence on the static structures,
dynamic properties and stability of ultracold quantum gases. Recent
experimental progress on dipolar BECs of chromium \cite{Griesmaier},
dysprosium \cite{Lu} and erbium \cite{Aikawa} atoms with large magnetic
dipole moments provides a unique opportunity to explore novel quantum phases
and quantum effects \cite{Lahaye}. Previous studies have shown that the
spontaneous breaking of both gauge symmetry and translational symmetry can
yield superfluidity and crystal periodic order in the dipolar BECs \cite%
{Josserand,Ancilotto1}. However, the attractive component of the DDI tends
to destabilize the condensate, leading to the system developing towards
collapse. It is possible to stabilize the system by taking account of
Lee-Huang-Yang (LHY) correction which is first-order correction to the mean
field energy $\left( \thicksim n^{5/2}\right) $ \cite{Huang}. In the
presence of LHY correction induced by quantum fluctuations, a new quantum
phase, that is, quantum droplet(QD) phase, is expected to appear. The QDs in
ultracold quantum gases of bosonic atoms are different from those in
superfluid helium\cite{Toennies,Barranco}, where the former have a larger
volume (two orders of magnitude) and a more dilute density (eight orders of
magnitude). Recently, stable QDs in dipolar BECs of dysprosium \cite{Kadau}
or erbium \cite{Chomaz1} atoms have been observed experimentally. These
experiments demonstrate that the quantum fluctuations can effectively
balance the attraction caused by the DDI and lead to the generation of
droplets, which has attracted much attention and interest. Afterwards, the
QDs in dipolar BECs were extensively investigated \cite%
{Ferrier-Barbut1,Baillie,Rafa,Abdel,Young} and opened up a new way for the
study of supersolids \cite%
{Modugno1,Langen,ChomazL,Modugno2,Roccuzzo,Stringari}. For instance, Chomaz $%
et$ $al.$ realized a long-lived supersolid state in dipolar BECs, and
observed the coexistence of density modulation and global phase coherence
\cite{ChomazL}. In addition, the quantum droplet states were also predicted
or observed in various cold atoms systems, such as binary boson mixtures
\cite{Petrov,Cabrera,Mishra,Smith,Dong,Elhadj,ZhangY,DuX}, heteronuclear
mixtures \cite{LSantos}, Bose-Fermi mixtures \cite{Rakshit}, spinor BECs
\cite{Ma,Oktel}, and mixtures of bosonic atoms and molecules with $p$-wave
interaction \cite{LiZ}. However, whether there exist QDs and phase transition from superfluid phase to supersolid phase in dipolar spin-1/2 BECs with attractive interspecies interactions and LHY correction is still unknown and deserves further investigation. In particular, how to effectively determine the supersolid phase in dipolar spinor BECs is an intriguing issue.

On the other hand, the spin-orbit coupling (SOC) is an interaction between
the spin and momentum of a quantum particle, which is ubiquitous in physical
systems \cite{Spielman}. The spin-orbit-coupled BECs provide us a brand-new
platform to investigate the novel quantum phenomena and exotic states of
matter due to its good experimental controllability. In the past few years,
cold-atom experiments have been successfully realized synthetic
one-dimensional (1D) \cite{Spielman}, 2D \cite{Wu} and 3D \cite{WangZY}
types of SOC, which are crucial in studying the exotic properties of
different dimensional topological matters. In addition, based on the
symmetry of interaction, there are three typical types of 2D SOCs in systems
including Rashba SOC \cite{Rashba}, Dresselhaus SOC \cite{Dresselhaus}, and
Rashba--Dresselhaus type SOC \cite{Goldman}. The SOC destroys the spatial symmetry and Galilean invariance, thus affecting the structure and superfluid properties of the condensates \cite{Zhai,Sakaguchi}. In this context, the competition between the SOC
and the contact interaction can lead to rich
quantum phases, topological structures and distinctive physical properties, such as plane-wave phase
\cite{Zhai,YZhang}, heliciform-stripe phase \cite{WangH}, topological
superfluid phase \cite{Wu}, supersolid phase \cite{Ketterle}, lattice
phase \cite{Sinha}, checkerboard phase \cite{Hu}, half-quantum vortex \cite%
{XuXQ}, soliton excitation \cite{Achilleos,Xu,ZhangYC,Kartashov1}, quantum beating \cite%
{WangQ}, and skyrmion \cite{Ramachandhran,LiuCF}, etc. Furthermore, the
effect of SOC on QDs has recently attracted considerable attention. In
ultracold Bose-Fermi mixtures, it was showed that SOC could independently
promote the formation of QDs \cite{Xiaoling}. And multi-dimensional solitons
and QDs were found in spin-orbit coupled binary BECs with LHY correction
\cite{Yongyao,XuSL}. To the best of our knowledge, however, there is seldom research on spinor (multi-component) BECs that simultaneously contain DDI, SOC and LHY correction. Such a physical system can be achieved under the current cold-atom experimental techniques \cite{PfauT}.

In this paper, we consider quasi-2D dipolar spin-1/2 BEC with quantum
fluctuations and Rashba SOC in a harmonic trap. The effects of LHY
corrections, DDI, SOC and rotation on the ground-state properties of the
system are analyzed. A ground-state phase diagram is presented as a function of DDI
strength and LHY correction strength in the absence of SOC. For given interspecies interaction
strengths, we can obtain the formation region of the droplet phase (the
equilibrium region of repulsion and attraction). In addition, we calculate the average energy of per atom for different quantum phases, which is helpful to further understand the changing process between different quantum phases of the system. Furthermore, in order to explore the supersolid properties of the system induced by LHY correction and DDI, we analyze the structural phase transitions of superfluid and supersolid
exhibited in the rotating case by investigating the ratio of the moment of inertia
to the rigid value and the nonclassical rotational inertia fraction. In the presence of SOC, the droplets become evidently
stretched and hidden vortex-antivortex pairs or clusters appear in the system. For a
non-droplet phase, the inclusion of SOC tends to make the system form a
droplet phase. Moreover, the system displays rich exotic spin textures
and topological structures including composite skyrmion-antiskyrmion-meron-antimeron cluster, meron(half-skyrmion)-antimeron(half-antiskyrmion)
string cluster, complicated antimeron-meron-antimeron chain cluster, and skyrmion-antiskyrmion-meron-antimeron necklace with a meron-antimeron necklace
embedded inside and a central spin Neel domain wall.

The paper is organized as follows. In Section 2, we formulate the theoretical model and methods. In Section 3, the ground-state structures are predicted and analyzed. In the absence of SOC, the effects of LHY corrections and DDI on the ground-state properties of the system are revealed, and the structural phase transition from superfluid to supersolid is demonstrated in terms of the moment of inertia and the nonclassical rotational inertia fraction. In the presence of SOC, the effect of SOC and the typical spin textures of the system are unveiled. In Section 4, the main conclusions of the paper are summarized.
\section{Theoretical model}
We consider a quasi-2D system of Rashba-type spin-orbit-coupled dipolar BECs
with LHY corrections in a harmonic trap. The dynamics of the system can be
described by the following nonlinear coupled Gross-Pitaevskii (GP) equations
\cite{Lahaye,Huang,Baillie,Petrov,Spielman,Zhai,Kartashov} {\small
\begin{align}
& i\hbar \frac{\partial \Psi _{1}}{\partial t} =\left( -\frac{\hbar ^{2}}{2m}
\nabla ^{2}+V(\bm{r})+g_{11}|\Psi _{1}|^{2}-g_{12}|\Psi _{2}|^{2}\right)
\Psi _{1}  \notag \\
&+\left(C_{dd}^{11}\int U_{dd}|\Psi _{1}(\bm{r}^{\prime},t)|^{2}\mathrm{d}%
\bm{r}^{\prime}+C_{dd}^{12}\int U_{dd}|\Psi _{2}(\bm{r}^{\prime},t)|^{2}%
\mathrm{d}\bm{r}^{\prime}\right) \Psi _{1}  \notag \\
&+G_{LHY}|\Psi _{1}|^{3}\Psi _{1}+\hbar \left( \kappa_{x}\partial
_{x}-i\kappa _{y}\partial _{y}\right) \Psi _{2},  \label{1} \\
& i\hbar \frac{\partial \Psi _{2}}{\partial t} =\left( -\frac{\hbar ^{2}}{2m}
\nabla ^{2}+V(\bm{r})+g_{22}|\Psi _{2}|^{2}-g_{21}|\Psi _{1}|^{2}\right)
\Psi _{2}  \notag \\
&+\left(C_{dd}^{21}\int U_{dd}|\Psi _{1}(\bm{r}^{\prime},t)|^{2}\mathrm{d}%
\bm{r}^{\prime}+C_{dd}^{22}\int U_{dd}|\Psi _{2}(\bm{r}^{\prime},t)|^{2}%
\mathrm{d}\bm{r}^{\prime}\right) \Psi _{2}  \notag \\
&+G_{LHY}|\Psi _{2}|^{3}\Psi _{2}-\hbar \left( \kappa_{x}\partial
_{x}+i\kappa _{y}\partial _{y}\right) \Psi _{1},  \label{2}
\end{align}%
}
where $m$ is the atomic mass, $\Psi _{j}\left( j=1,2\right) $ is the
component wave function, with 1 and 2 corresponding to spin-up and
spin-down, respectively. $V(\bm{r})=\frac{1}{2}m[\omega _{\bot
}^{2}(x^{2}+y^{2})+\omega _{z}^{2}z^{2}]$ is the external trapping
potential. The coefficients $g_{jj}=4\pi a_{j}\hbar ^{2}/m$ $(j=1,2)$ and $%
g_{12}=g_{21}=4\pi a_{12}\hbar ^{2}/m$ represent the intra- and interspecies
coupling strengths, where $a_{j}$ $(j=1,2)$ and $a_{12}$ denote the $s$-wave
scattering lengths between intra- and intercomponent atoms. Here we consider
$a_{11}=a_{22}=a_{s}$, i.e., the two component atoms have the same
intraspecies $s$-wave scattering lengths. $\kappa _{x}$ and $\kappa_{y}$
characterize the Rashba SOC strengths in the $x$ and $y$ directions.

The long-range nonlocal DDI can be expressed as \cite{Lahaye}
\begin{equation}
U_{dd}\left( \bm{r}-\bm{r}^{\prime }\right) =\frac{1-3\cos ^{2}\theta }{%
\left\vert \bm{r}-\bm{r}^{\prime }\right\vert ^{3}},  \label{3}
\end{equation}%
with $\theta $ being the angle between the polarization direction and the
relative position of the atoms. $C_{dd}^{jj}=\mu _{0}\mu _{j}^{2}/4\pi $ $%
(j=1,2)$ and $C_{dd}^{12}=C_{dd}^{21}=\mu _{0}\mu _{1}\mu _{2}/4\pi $ are
the magnetic DDI constants of intraspecies and interspecies components,
respectively. $\mu _{0}$ denotes the permeability of vacuum, and $\mu _{j}$ $%
(j=1,2)$ is the magnetic dipole moment of the $j$th component atom. Here $%
\mu _{1}=\mu _{2}=\mu $ is assumed, which means $%
C_{dd}^{11}=C_{dd}^{22}=C_{dd}^{12}=C_{dd}^{21}=C_{dd}=\mu _{0}\mu ^{2}/4\pi
$. The 3D dipolar length is defined as $a_{dd}=\mu _{0}\mu _{j}^{2}m/12\pi
\hbar ^{2}$ and the ratio of DDI to the $s$-wave interaction strength is
given by $\varepsilon _{dd}=a_{dd}/a_{s}$.

In Eq.(\ref{1}) and Eq.(\ref{2}), quantum fluctuations for the mean-field
energy of the system are introduced by a LHY correction with coefficient $%
G_{LHY}=\frac{128\sqrt{\pi }\hbar ^{2}a_{s}^{5/2}}{3m}\left( 1+\frac{3}{2}%
\varepsilon _{dd}^{2}\right) $ \cite{Cidrim}. In the present work, we assume
that the strength of SOC is relatively weak and does not affect the form of
the LHY correction \cite{Yongyao,Sachdeva}. Meanwhile, we omit the direct
effect of intercomponent interaction on the LHY correction, but take into
account its indirect influence on the LHY correction, as shown in the
coupled nonlinear GP equations of the binary BECs. This treatment does not
lead to essential changes in the physical properties of the system because
in general the effect of LHY correction on the system is primarily
attributed to the quartic nonlinearities in form \cite{Petrov,ZhangY,Dalfovo}%
. Another reason for our choice of this approach is that the analytical
solution and numerical calculation of the LHY correction term for
spin-orbit coupled dipolar BEC mixtures become rather difficult due to the
complexity of the derivation process and calculation formula for the strict
LHY expression of the present system. As a matter of fact, a simplified
treatment scheme similar to our method has been successfully applied to the
investigation of vortical droplets in two-component swirling superfluids
\cite{Kartashov}. Here we are interested in a quasi-2D system with strong
confinement in the $z $ direction. In this case, we separate the degrees of
freedom of the wave function as $\Psi _{1,2}(\bm{r},t)=\psi
_{1,2}(x,y,t)\phi _{1,2}(z)$, and integrate the $z$ dependence. $\phi
_{1,2}(z)=( \frac{1}{\sqrt{\pi }a_{z}}) ^{1/2}\exp ( \frac{%
-z^{2}}{2a_{z}^{2}}) $ denotes the single-particle ground-state wave
function in a harmonic potential with $a_{z}=\sqrt{\hbar /m\omega _{z}}$.
The normalization condition of the system reads $\int \left[ |\psi
_{1}|^{2}+|\psi _{2}|^{2}\right] \mathrm{d}x\mathrm{d}y=N$, where $N$ is the
number of atoms.

In order to perform numerical calculation and simulation, we introduce the
notations $\widetilde{t}=\omega _{\perp }t$, $\widetilde{x}=x/a_{0}$, $%
\widetilde{y}=y/a_{0}$, $\widetilde{V}(x,y)=V(x,y)/\hbar \omega _{\perp }$, $%
\widetilde{\psi }_{j}=\psi _{j}a_{0}/\sqrt{N}(j=1,2)$, and $a_{0}=\sqrt{%
\hbar /m\omega _{\bot }}$. Then we obtain the dimensionless
2D coupled GP equations
\begin{align}
& i\partial _{t}\psi _{1} =\left( -\frac{1}{2}\nabla ^{2}+V+\beta _{11}|\psi
_{1}|^{2}-\beta _{12}|\psi _{2}|^{2}\right) \psi _{1}  \notag \\
&+c_{dd}\mathscr{F}_{2D}^{-1}\left[ \widetilde{n}(\bm{k,}t)F(\bm{k}a_{z}/%
\sqrt{2})\right] \psi _{1}+g_{LHY}|\psi_{1}|^{3}\psi _{1}  \notag \\
&+\left( \kappa _{x}\partial _{x}-i\kappa_{y}\partial _{y}\right) \psi _{2},
\label{4} \\
& i\partial _{t}\psi _{2} =\left( -\frac{1}{2}\nabla ^{2}+V+\beta _{22}|\psi
_{2}|^{2}-\beta _{21}|\psi _{1}|^{2}\right) \psi _{2}  \notag \\
&+c_{dd}\mathscr{F}_{2D}^{-1}\left[ \widetilde{n}(\bm{k,}t)F(\bm{k}a_{z}/%
\sqrt{2})\right] \psi _{2}+g_{LHY}|\psi_{2}|^{3}\psi _{2}  \notag \\
&-\left( \kappa _{x}\partial _{x}+i\kappa _{y}\partial _{y}\right) \psi _{1},
\label{5}
\end{align}
where the tildes are omitted for simplicity in our following discussion. In
other words, for the convenience of discussion, we retain the same symbols
as the dimensional variables unless otherwise indicated. Here $\beta
_{11}=\beta _{22}=2\sqrt{2\pi }a_{s}N/a_{z}$ and $\beta _{12}=\beta _{21}=2%
\sqrt{2\pi }a_{12}N/a_{z}$ are the dimensionless intra- and intercomponent
interaction strengths. The 2D dimensionless LHY correction coefficient is
given by $g_{LHY}=\frac{128\sqrt{2}a_{s}^{5/2}N^{3/2}}{3\sqrt{5}\pi
^{1/4}a_{0}a_{z}^{3/2}}\left( 1+\frac{3}{2}\varepsilon _{dd}^{2}\right)$. $%
c_{dd}=\mu _{0}\mu ^{2}mN/\left( 3\sqrt{2\pi }\hbar ^{2}a_{z}\right) $
represents the dipolar coupling efficient, and $\mathscr{F}_{2D}$ denotes
the 2D Fourier transform operator with $\widetilde{n}(\bm{k},t)=\mathscr{F}%
_{2D}[n(\bm{r},t)]$ \cite{Ticknor}. The function $F(\bm{q})$ with $%
\bm{q\equiv k}a_{z}/\sqrt{2}$ is the $k$-space DDI for the quasi-2D
geometry, which consists of two parts, originating from polarization
perpendicular or parallel to the direction of the dipole tilt. More
specifically, $F(\bm{q})=\cos ^{2}(\alpha )F_{\perp }(\bm{q})+\sin
^{2}(\alpha )F_{\parallel }(\bm{q})$ with $\alpha $ being the angle between
the $z$-axis and the polarization vector $\hat{d}$. Here $F_{\perp }(\bm{q}%
)=2-3\sqrt{\pi }qe^{q^{2}}$erfc$(q)$ ($\bot $-configuration), $F_{\parallel
}(\bm{q})=-1+3\sqrt{\pi }(q_{d}^{2}/q)e^{q^{2}}$erfc$(q)$ ($\Vert $%
-configuration), $q_{d}$ is the wave vector along the direction of the
projection of $\hat{d}$ onto the $x$-$y$ plane, and erfc denotes the
complementary error function \cite{Fischer,Nath}. We consider that the
polarization is vertical to the condensate plane, i.e., $\alpha =0$, and
therefore $F_{\perp }(\bm{q})=2-3\sqrt{\pi }qe^{q^{2}}$erfc$(q)$.

In order to describe the topological properties of the system, we use a
nonlinear Sigma model \cite{Aftalion,Kasamatsu2}, in which a normalized
complex-valued spinor $\bm{\chi }=[\chi _{1},\chi _{2}]^{T}$ with $|\chi
_{1}|^{2}+|\chi _{2}|^{2}=1$ is introduced. The corresponding two-component
wave function can be written as $\psi _{1}=\sqrt{\rho }\chi _{1}$ and $\psi
_{2}=\sqrt{\rho }\chi _{2}$, where $\rho =|\psi _{1}|^{2}+|\psi _{2}|^{2}$
is the total density of the system. The spin density is given by $\bm{S}=%
\overline{\bm{\chi }}\bm{\sigma}\bm{\chi}$, where $\bm{\sigma}=(\sigma
_{x},\sigma _{y},\sigma _{z})$ are the Pauli matrices, and the components of
$\bm{S}$ are expressed as
\begin{eqnarray}
S_{x} &=&\chi _{1}^{\ast }\chi _{2}+\chi _{2}^{\ast }\chi _{1},  \label{Sx}
\\
S_{y} &=&i(\chi _{2}^{\ast }\chi _{1}-\chi _{1}^{\ast }\chi _{2}),
\label{Sy} \\
S_{z} &=&|\chi _{1}|^{2}-|\chi _{2}|^{2},  \label{Sz}
\end{eqnarray}%
with $|\bm{S}|^{2}=S_{x}^{2}+S_{y}^{2}+S_{z}^{2}=1$. The spacial
distribution of the topological structure of the system is described by the
topological charge density%
\begin{equation}
q(x,y)=\frac{1}{4\pi }\bm{S}\cdot \left( \frac{\partial \bm{S}}{\partial x}%
\times \frac{\partial \bm{S}}{\partial y}\right) ,
\label{TopologicalChargeDensity}
\end{equation}%
and the topological charge $Q$ is defined as
\begin{equation}
Q=\int q(x,y)\mathrm{d}x\mathrm{d}y.  \label{TopologicalCharge}
\end{equation}

\section{Results and discussion}

For such a complex system, there is no analytical
solution. In what follows, we numerically solve the 2D nonlinear coupled GP equations (\ref{4})
and (\ref{5}) to obtain the ground state of the system by using the
imaginary-time propagation method \cite{YZhang,LiJ}. In our simulation, we consider two-component
BECs of $^{164}$Dy atoms and choose the parameters of the attractive
interspecies interactions as $\beta _{12}=\beta _{21}=100$. It is shown that
the system can exhibit rich and exotic quantum phases due to the competition
among multiple parameters.

\subsection{The effects of LHY corrections and DDI}

We first study the effects of LHY corrections and DDI on the ground-state
properties of the system without SOC in the nonrotating case. In Fig. 1 we present a ground-state phase diagram with respect to the DDI strength $c_{dd}$
and the LHY correction strength $g_{LHY}$. There are six different quantum phases
marked by A--F and a collapse region, which differs in terms of their density
profiles. For other interaction strengths, our simulation results show that
there are similar phase diagrams and ground-state configurations. In the
following discussion, we will give a detailed description for individual quantum phases. Due to the attractive interspecies interaction, the two-component density distributions are the same, so we only need to plot the density profile of component 1. The typical density distributions of the six different phases A--F in Fig. 1 are shown
in the first row of Figs. 2(a)-2(f), respectively. The corresponding momentum distributions, i.e., the $k$-space densities of component 1 are displayed in the second row of Fig. 2.

\begin{figure}[htbp]
\centerline{\includegraphics*[width=7cm]{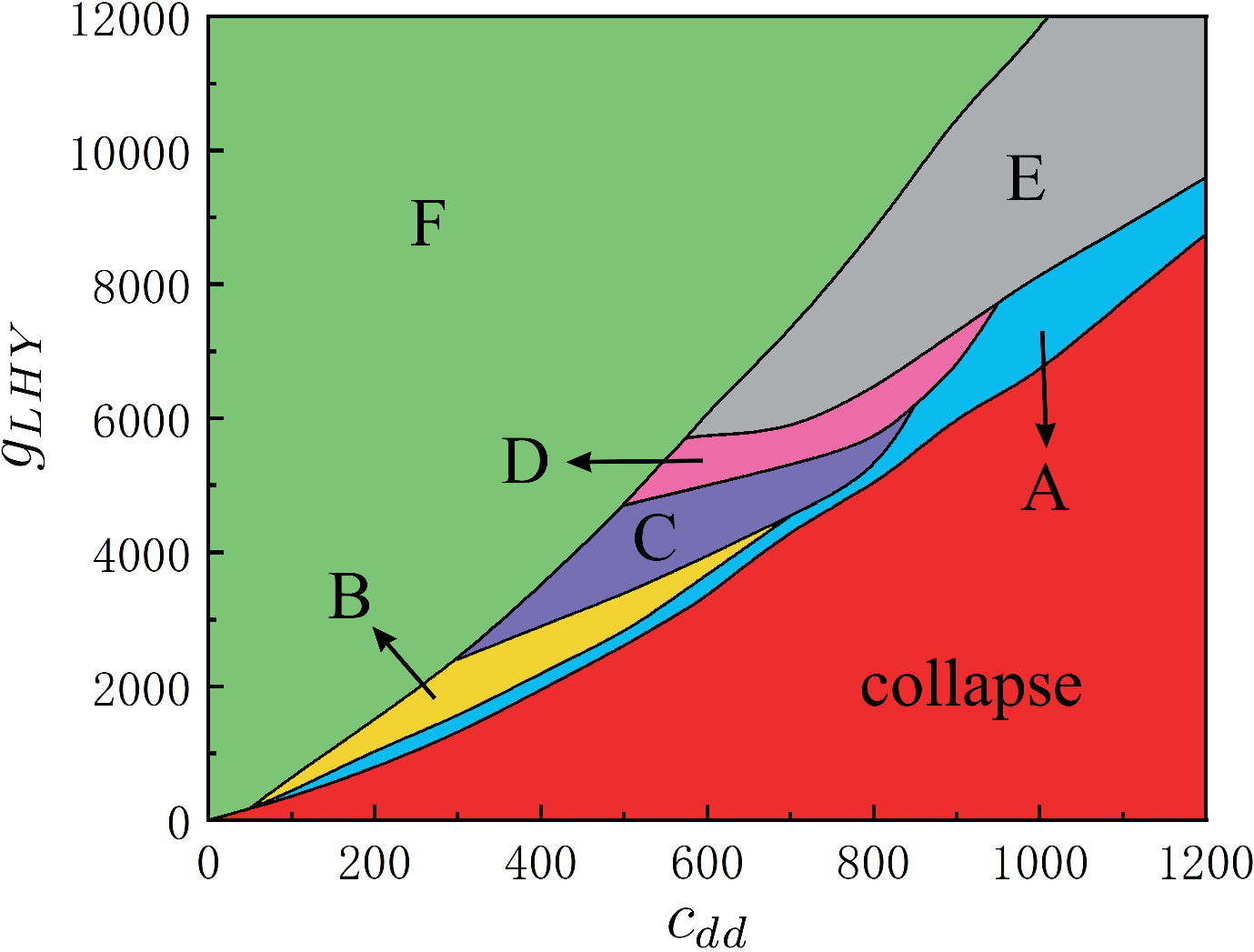}}
\caption{(Color online) Ground-state phase diagram spanned by the DDI strength $c_{dd}$ and the LHY
correction strength $g_{LHY}$ for two-component dipolar BECs with quantum fluctuations, where $\beta _{12}=\beta _{21}=100$. There are six different quantum phases marked by A-F and a
collapse regime.}
\label{Figure1}
\end{figure}

The phase diagram includes seven different regions. The bottom red region in Fig. 1 represents system collapse, where the attractive force originated from the interspecies attractive interaction and the DDI is dominant, and the repulsive force is not strong enough to stabilize the system \cite{Mardonov}. Recently, relevant research demonstrated that the LHY quartic term induced by quantum fluctuations was sufficient for the stabilization of ultracold 2D Bose gas against the collapse \cite{Shamriz}. The other six regions denote six different ground-state quantum phases labelled by A--F. For a fixed DDI strength within a large range, when the LHY correction strength increases, the system sustains A phase which is displayed by the region A in Fig. 1. In this phase, the density distribution shows an evident quantum droplet lattice (see Fig. 2(a)). At the same time, the momentum distribution is concentrated at several discrete
points, where the central high density point is surrounded by some low density points, which indicates that the atoms are condensed in discrete narrow momentum regimes with very finite momentums. In addition, the momentum distribution exhibits excellent crystal order with axial symmetry concerning the $k_{x}$ axis and the $k_{y}$ axis. We may call the A phase in Fig. 1 as droplet lattice phase. Physically, sufficiently large repulsive force stemming from the LHY corrections of the system is the key factor for the droplets to be in a stable state, which is balanced with the attractive force induced by the DDI and intercomponent attractive interaction. For fixed contact interaction strengths, there is a significant competitive relationship between the LHY correction strength and the DDI strength on the ground-state structure of the system.

\begin{figure}[htbp]
\centerline{\includegraphics*[width=8cm]{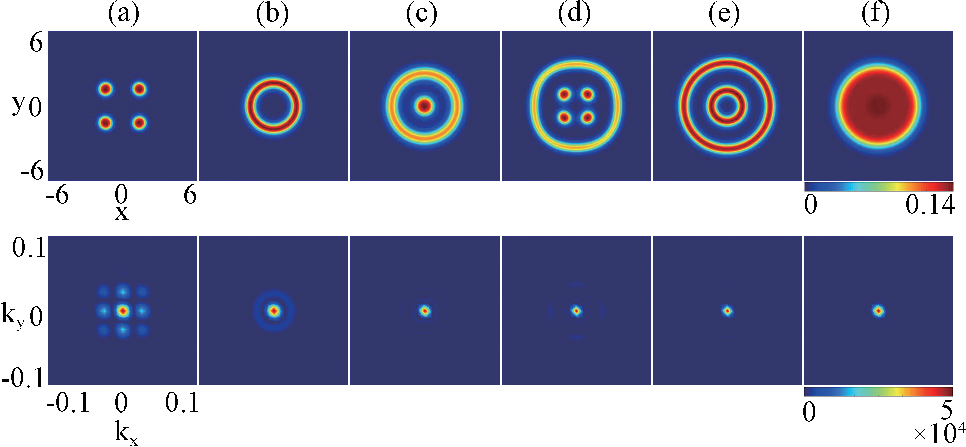}}
\caption{(Color online) Ground-state quantum phases of two-component dipolar BECs with LHY
corrections, where (a)-(f) correspond to the A-F phases in Fig.1, respectively.
The first row denotes the density distribution of component 1, and the second row corresponds to the momentum distribution of component 1. (a) $c_{dd}=300$, $g_{LHY}=1300$, (b) $c_{dd}=300$, $g_{LHY}=1600$, (c) $c_{dd}=500$, $g_{LHY}=3700$, (d) $c_{dd}=900$, $g_{LHY}=7200$, (e) $c_{dd}=900$, $g_{LHY}=7700$, and (f) $c_{dd}=300$, $g_{LHY}=3300$. The unit
length is $a_{0}$.}
\label{Figure2}
\end{figure}

For relatively weak DDI, when the LHY correction strength slightly increases, the A phase transforms to the B phase as shown in Fig. 1 and Fig. 2(b). The B phase is an annular condensate (annular phase), where the density distribution of the system exhibits a typical annular structure, and the momentum distribution is concentrated at a central point surrounded by a low-density ring (Fig. 2(b)). For moderate DDI intensity, with the increase of LHY correction strength, the C phase or the D phase emerges as the ground state of the system, which is displayed in Fig. 1 and Figs. 2(c)-2(d). The C phase is a droplet-stripe phase, where the central quantum droplet is surrounded by a peripheral toroidal stripe (see Fig. 2(c)). Intriguingly, the D phase denotes an exotic coexistence phase of a droplet lattice and an annular stripe, where the central region embraced by a circular density stripe holds a square droplet lattice (Fig. 2(d)). From the momentum distributions in Figs. 2(c)-2(d), there is only a discrete point near the origin of momentum space, which implies that the atoms are essentially condensed at zero momentum. In particular, the coexistence phase of droplet lattice and annular stripe (the D phase) has not been reported elsewhere so far, and it allows to be tested and observed in the future cold atom experiments. In the case of strong DDI, with the increase of LHY correction strength, the A phase transforms to the E phase as indicated by the region E in Fig.1. The typical density and momentum distributions of this phase given in Fig. 2(e), where the ground-state density forms two concentric layered annular stripes. We can call the E phase as annular stripe phase. When the repulsive force induced by LHY correction is much greater than the attractive force caused by DDI, the system supports typical Thomas-Fermi (TF) phase (plane-wave phase) as shown in Fig. 1 and Fig. 2(f). Recently, the TF phase has also been observed in nonrotating spin-orbit coupled BECs \cite{Zhai,YZhang,WangH}, where the formation of this phase is mainly attributed to the competition between the SOC and the interatomic contact interaction. From Fig.1, one can see that, for an appropriate regime of the DDI strength, the phase transitions of B$\rightarrow$C$\rightarrow$D$\rightarrow$E$\rightarrow$F phases can be achieved by varying the LHY correction strength. In addition, for large DDI strength, the collapse area in the phase diagram significantly expands. The physical reason is that the system needs sufficient repulsion to maintain stability under the strong attraction caused by the DDI. Particularly, it is expected that a supersolid phase can be formed in the system due to the existence of QDs with high atomic density (e.g., see the A, C and D phases), and the finite nonclassical rotational inertia of the system \cite{Roccuzzo} will be examined in the following sections to further discuss the properties of the system.

\begin{figure}[htbp]
\centerline{\includegraphics*[width=7cm]{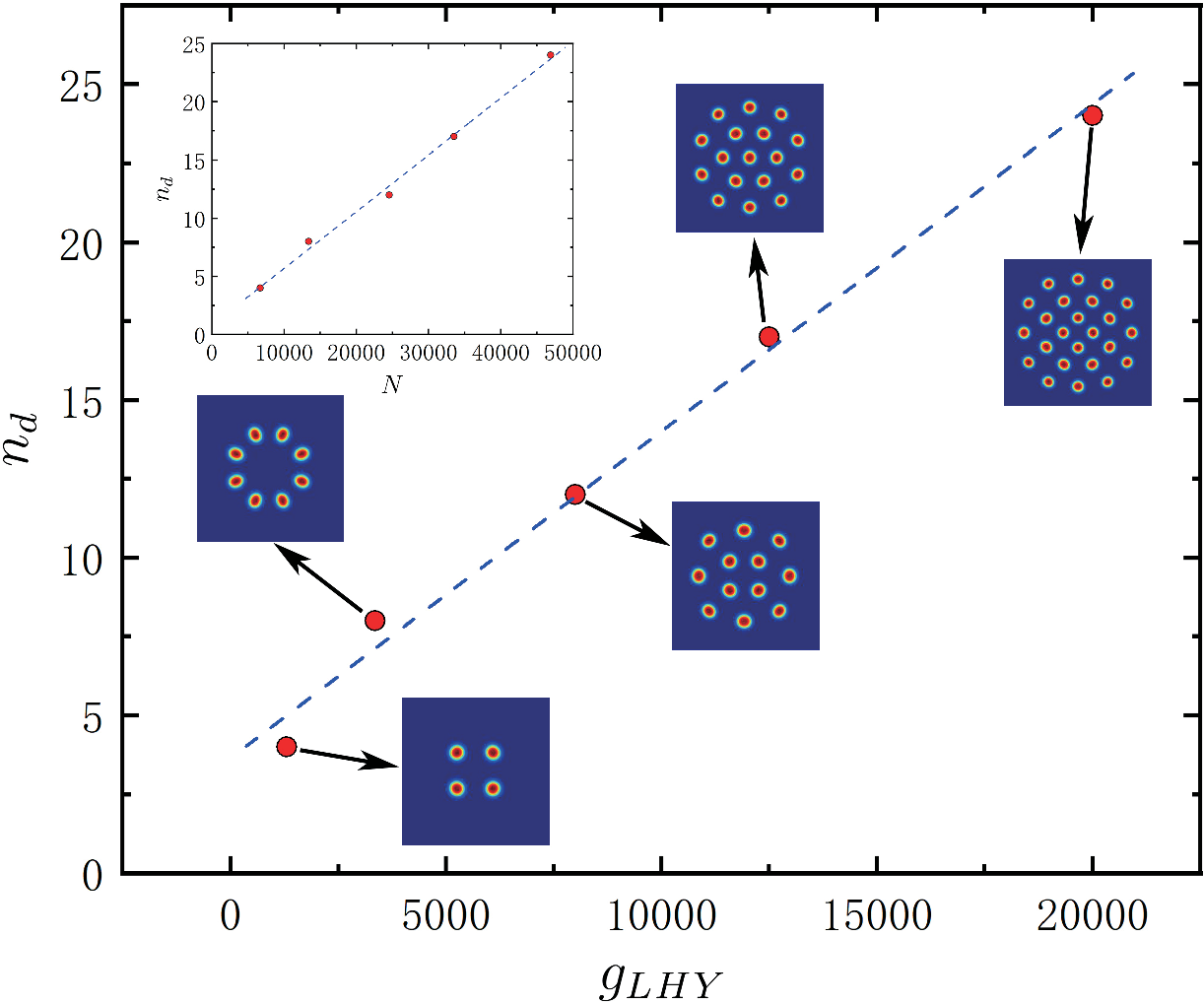}}
\caption{(Color online) The number of droplets $n_{d}$ with respect to the LHY
correction coefficient $g_{LHY}$. The parameters of the five marked red dots from left to right are (1) $%
c_{dd}=300, g_{LHY}=1300 $, (2) $c_{dd}=600, g_{LHY}=3350 $, (3) $%
c_{dd}=1100, g_{LHY}=8000$, (4) $c_{dd}=1500, g_{LHY}=12500$, and (5) $%
c_{dd}=2100, g_{LHY}=20000$, respectively. The five insets pointed by the arrows represent the density distributions at these five different dots. At the same time, the inset in the upper left corner shows the number of droplets $n_{d}$ as a function of the number of atoms $N$, corresponding to the five different dots mentioned above. The blue dashed lines are fitted linear curves.}
\label{Figure3}
\end{figure}

With regard to the droplet lattice phase (the A phase), we now study the relation between the number of droplets $n_{d}$ and the LHY correction strength $%
g_{LHY}$, as shown in Fig. 3. As $c_{dd}$ and $g_{LHY}$ increase together (this is equivalent to an increase of the number of atoms $%
N$), the number of droplets $n_{d}$ gradually increases and shows an approximate linear dependence on the LHY correction coefficient $g_{LHY}$. Obviously, increasing the LHY correction strength or increasing the number of atoms results in growth of the microscopic droplet lattice. Physically, with the increase of LHY correction strength, or equivalently, with the increase of the number of atoms, the original droplets tend to redistribute and reorganize into more droplets to decrease the system energy and the confinement energy in the $z$ direction. Recently, the linear dependence between the number of droplets and the number of atoms has been observed and verified in the quantum ferrofluid experiment of dysprosium atomic BEC with DDI and quantum fluctuations \cite{Kadau}. Our numerical simulation shows that no QDs are generated in the system when the number of atoms is less than a certain critical value. The main reason is that the effective repulsive force resulted from the LHY corrections is too weak to meet the formation condition of QD. In addition, the droplets exhibit regular patterns (such as square droplet lattice, droplet necklace and triangular droplet lattice) reflecting the symmetry of the system. Furthermore, the linear relation between the number of droplets and the number of atoms in Fig. 3 is consistent with recent experimental observations \cite%
{Kadau}, which indirectly demonstrates the rationality and validity of our theoretical model.

\begin{figure}[tbph]
\centerline{\includegraphics*[width=8cm]{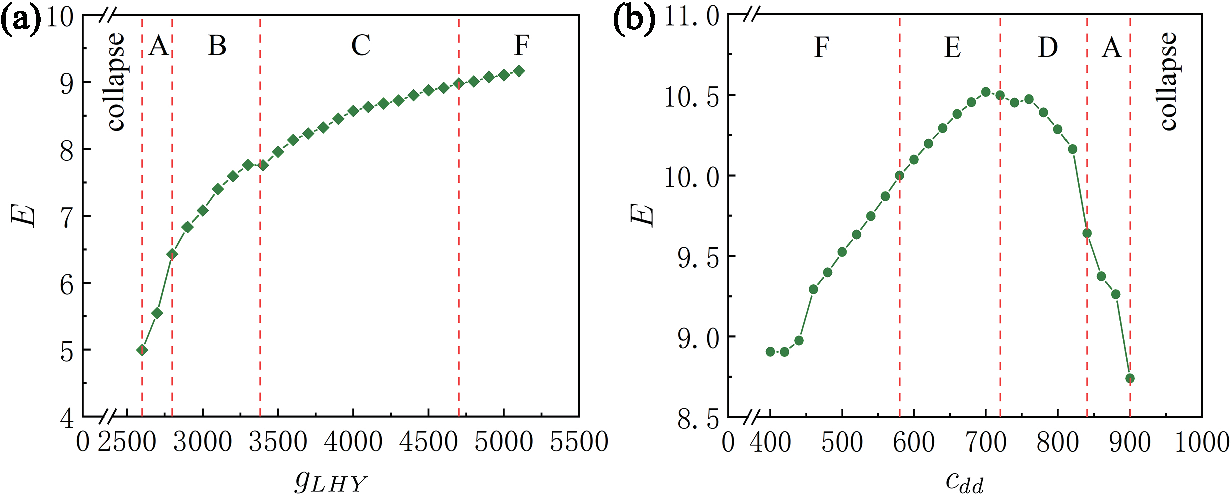}}
\caption{(Color online) Energy per atom $E$ as a
function of (a) $g_{LHY}$ for $c_{dd}=500$ and (b) $c_{dd}$ for $g_{LHY}=6000$. The vertical dotted lines separate different phases.}
\label{Figure4}
\end{figure}

For a given set of parameter values, the ground state of the system corresponds to the minimum of system energy. The energy per atom for a stationary state in the absence of SOC is given
by
\begin{align}
E=&\int \mathrm{d}x\mathrm{d}y\Big[\sum_{j=1,2}\psi _{j}^{\ast }\Big(-%
\frac{1}{2}\nabla ^{2}+V+\frac{1}{2}\Phi   \notag \\
& +\frac{2}{5}g_{LHY}|\psi _{j}|^{3}\Big)\psi _{j}+\frac{1}{2}\beta
_{11}|\psi _{1}|^{4}+\frac{1}{2}\beta _{22}|\psi _{2}|^{4}  \notag \\
& -\beta _{12}|\psi _{1}|^{2}|\psi _{2}|^{2}\Big],  \label{energy1}
\end{align}%
with $\Phi =c_{dd}\mathscr{F}_{2D}^{-1}\left[ \widetilde{n}(\bm{k,}t)F(\bm{k}%
a_{z}/\sqrt{2})\right] $. Fig. 4(a) and Fig. 4(b) show the energy per atom as a
function of $g_{LHY}$ for $c_{dd}=500$ and that of $c_{dd}$ for $g_{LHY}=6000$, respectively, where A-F and collapse correspond to the different quantum phases and the system collapse region in Fig. 1. In the stable quantum phase region, as the LHY correction strength increases, the energy per atom tends to increase overall (Fig. 4(a)), and finally the increasing inclination slows down. The reason is that the increase of the LHY correction strength enhances the effective repulsive energy of atoms. In Fig. 4(b), the energy per atom gradually increases with the increasing of $c_{dd}$ for the F and E phases, while it non-monotonically decreases with the increasing of $c_{dd}$ for the D and A phases. This feature can be understood. In the F and E phases, the effective two-body interactions in each component are repulsive as the DDI is tuned positive by the oblate TF density distribution and the toroidal stripe density profile of the condensate (e.g., see Fig. 2(f) and Fig. 2(e)). In other words, the oblate TF density distribution and the toroidal stripe density profile boost the repulsive side-by-side interaction of the magnetic dipoles \cite{Baillie,Young,Bisset}. By comparison, in the D and A phases, the effective two-body interactions in each component become attractive in view of the prolate shape of the droplets for the 3D case \cite{Baillie,Young}, where the $z$ confinement becomes important as $c_{dd}$ further increases. Moreover, the energy conversion process and angular momentum of rotating superfluid droplets can well reflect the structural characteristics of the system \cite{Sean}, which is of great significance for us to further study the properties of superfluids and supersolids.

\subsection{Moment of inertia}

Next, we investigate the moment of inertia by adding the term $-\Omega \widehat{%
L}_{z}\psi $ to Eq.(\ref{4}) and Eq.(\ref{5}) and
calculating the angular momentum. Without loss of generality, we introduce a relatively small rotation
frequency $\Omega =0.5$. The moment of inertia per atom $\Theta$ relative to the $z$ axis can be defined through the relationship $\left\langle
L_{z}\right\rangle =\Omega\Theta$. Then one can get $\Theta$ by calculating the
mean angular momentum per atom. If the moment of inertia becomes large
and even approaches the rigid value, $\Theta _{rig}=\int \mathrm{d}x\mathrm{d}%
y\left( x^{2}+y^{2}\right) n\left( x,y\right) $, we can show that the
density profile is not rotationally invariant as a consequence of the
mechanical drag caused by the rotation \cite{Roccuzzo}. In the meantime, the moment
of inertia fixes the nonclassical rotational inertia (NCRI) fraction $%
f_{NCRI}=1-\Theta /\Theta _{rig}$ in the isotropic harmonic potential with $%
\omega_{x}=\omega_{y}$ \cite{Modugno2,Roccuzzo}. The NCRI can be used as
a direct strong evidence to describe superfluidity under rotation. In Fig. 5, we
report our calculation results for the ratio of the moment of inertia per atom $\Theta$ and the rigid value $\Theta _{rig}$, and give the density and phase distributions corresponding
to different stages. When $g_{LHY}=5000$, we find that
the ratio $\Theta /\Theta _{rig}$ of the system jumps three times in Fig. 5(a). The first
jump occurs at $c_{dd}\approx 250$, which is due to the increase of vortex number and the transition of vortex structure (from vortex necklace to triangular vortex lattice) caused by the increase of DDI strength. The increase of the vortex number in the system leads to a significant enlargement of the average angular momentum per atom. The second transition occurs at $c_{dd}\approx 510$, where the triangular vortex lattice composed of seven vortices suddenly transforms into a compact vortex necklace comprised of five vortices with closer and larger vortex cores, and the ratio $\Theta /\Theta _{rig}$ evidently decreases. These jumps belong to the structural phase transitions within the superfluid. When the DDI strength
becomes larger, the ground state of the rotating system enters the supersolid phase
(corresponding to regions A, C and D in Fig.1 under the nonrotating case) as shown in Fig. 5(a), and we can see that the ratio $\Theta /\Theta _{rig}$ has an evident higher jump and is numerically close to 1, i.e., the moment of inertia per atom $\Theta$ approaches the rigid value $\Theta _{rig}$. Specifically, the ratio $\Theta /\Theta _{rig}$ significantly increases at $c_{dd}\approx590$, indicating a phase transition of the system from superfluid
phase to supersolid phase. Essentially, here the supersolid phase is a quantum droplet lattice phase containing hidden vortices \cite{WenL1,WenL2}, which possesses both superfluid and solid characteristics. From Fig. 5(b), it is shown that the corresponding NCRI fraction $f_{NCRI}$ and the ratio $\Theta /\Theta _{rig}$ maintain synchronous jumps, and $f_{NCRI}<1$ manifests the superfluid property of the system \cite%
{Modugno2}. There is a visible jump of physical quantity $f_{NCRI}$ at the transition boundary between the superfluid phase and
the supersolid phase, which demonstrates the first-order phase transition of the
system \cite{Stringari,Rica,Pohl,Shlyapnikov}. Obviously, the $f_{NCRI}$ reaches its
lowest value when the system is in the supersolid phase (Fig. 5(b)), where the system exhibits
both fluid and solid behaviors. In a recent cold-atom supersolid experiment, Modugno $%
et $ $al.$ observed that the supersolid phase in a single-component dipolar BEC
features a significantly increased moment of inertia relative to an ordinary scalar BEC \cite{Modugno2}. Therefore, our theoretical predictions are expected to be tested, confirmed, and further studied in the future dipolar supersolid experiments.
\begin{figure*}[tbph]
\centerline{\includegraphics*[width=14cm]{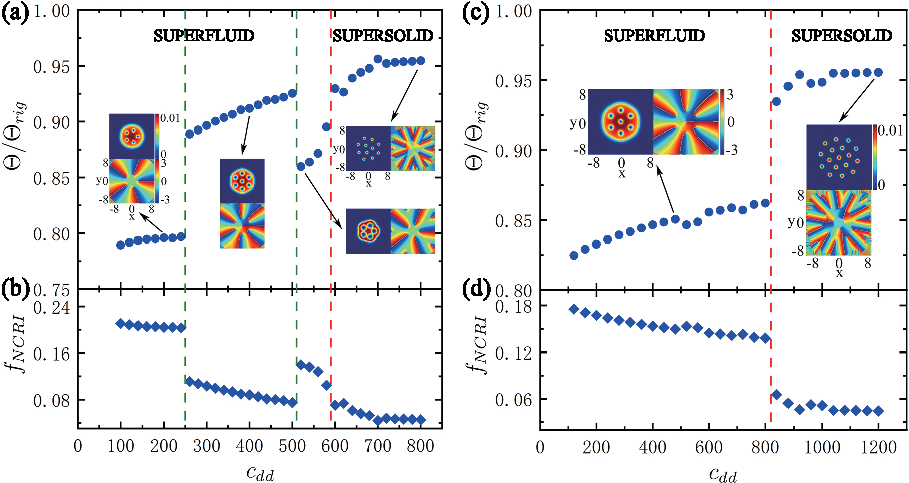}}
\caption{(Color online) (a) and (c) The ratio of the moment of inertia per atom $\Theta$ and the rigid value $\Theta _{rig}$ in rotating two-component dipolar BECs with LHY corrections, as a function of $c_{dd}$. The insets are the typical density distributions and phase distributions for different stages of the system. (a) $g_{LHY}=5000$, $\Omega =0.5$, and (c) $g_{LHY}=9000$, $\Omega =0.5$. (b) and (d) The corresponding nonclassical rotational inertia fraction as a function of $c_{dd}$.}
\label{Figure5}
\end{figure*}

For the case of $g_{LHY}=9000$, the ratio $\Theta /\Theta _{rig}$ and the NCRI fraction $f_{NCRI}$ versus $c_{dd}$ are shown in Fig.5(c) and Fig. 5(d), where the change trends are similar to those in Fig. 5(a) and Fig. 5(b), respectively. In this case, the first-order phase transition from superfluid phase to supersolid phase occurs at $c_{dd}\approx 820$, where $\Theta /\Theta _{rig}$ and $f_{NCRI}$ simultaneously undergo a sudden change. The configuration of the typical supersolid phase (see Fig. 5(c)) is also a quantum droplet lattice phase with hidden vortices (to be exactly, a quantum droplet lattice phase with a hidden vortex necklace plus a hidden triangular vortex lattice). As $c_{dd}$
increases, the moment of inertia eventually approaches a steady value close to the rigid value,
reflecting the crystalline nature of the QDs. Compared to the nonrotating case, for the supersolid phase, the distribution of droplets in the system (see Fig. 3 and Fig. 5(c)) is basically not broken under the action of rotation due to the partial rigid body nature of the droplets. In addition, each droplet in the supersolid phase is essentially itself superfluid, which is the reason why the rigid body value is not exactly achieved. Nevertheless, compared to the distance between droplets (Fig. 5(c)), the size of each droplet is so small that the difference between the moment of inertia per atom in the supersolid phase and the rigid body value can be almost negligible.

\subsection{The effect of SOC}

Then we study the effect of SOC on the ground-state structure of the
system in the absence of rotation, and the main results are illustrated in Fig. 6, where the first and fourth rows are the density distributions, the second and fifth rows denote the phase distributions, and the third and sixth rows represent the corresponding momentum distributions (i.e., the $k$-space densities of the system)). The SOC strength for the top three rows of Fig. 6 is $\protect\kappa _{x}=\protect\kappa _{y}=2$ and that for the bottom three rows is $\protect\kappa _{x}=\protect\kappa _{y}=6$. The relevant parameters in Figs. 6(a)-6(f) are the same as those in Figs. 2(a)-2(f). In this case, the system still exhibits a miscible phase, which is due to the fact that here the interspecies interactions are attractive while those in Ref. \cite{Gui} are repulsive, thus we only show the density profile of component 1. In the phase distributions, the value of the phase varies continuously from $-\pi $ to $\pi$, where the end point of the boundary between a $\pi$ phase line and a $-\pi$ phase line represents a quantum vortex (anticlockwise rotation) or antivortex (clockwise rotation). It is well known that there are three fundamental types of vortices in cold atom physics: visible vortex, ghost vortex, and hidden vortex \cite{LiJ,WenL1,WenL2,Fetter,Ueda}. The visible vortex is the conventional quantized vortex that is visible in both the density distribution and the phase profile and carries angular momentum \cite{WenL1,Fetter}. For the ghost vortex, it shows up in the phase distribution as a phase singularity but has no visible vortex core in the density distribution and carries no angular momentum \cite{Ueda}. In contrast, the hidden vortex is visible in the phase distribution but invisible in the density profile and it carries angular momentum \cite{LiJ,WenL1,WenL2}. Only after including the hidden vortices can the well-known Feynman rule be satisfied. Comparing Fig. 6 with Fig. 2, one can clearly see the influence of SOC on the quantum phases of the system. For relatively weak 2D SOC, e.g., $\kappa_{x}=\kappa _{y}=2$ (the top three rows), when $c_{dd}=300$ and $g_{LHY}=1300$ quantum droplets are stretched along the azimuth direction, while four hidden vortex-antivortex pairs are generated around the QDs (see Fig. 6(a1)). Obviously, the introduction of SOC breaks the spatial symmetry of the system, leading to the stretching of droplets. The momentum distribution shifts towards the negative direction of $k_{y}$ as a whole, and some low density points are stretched into narrow stripe shapes. Thus the ground state of the system becomes a square stretched droplet lattice phase with hidden vortex-antivortex pairs. Compared with Fig. 2(b), due to the SOC effect, the ground state evolves from an annular phase into a triangular droplet lattice phase with hidden vortex-antivortex cluster (Fig. 6(b1)). At the same time, the $k$-space density is principally concentrated at two discrete points, which means that the atoms are mainly condensed at two finite momenta.

\begin{figure}[tbph]
\centerline{\includegraphics*[width=8cm]{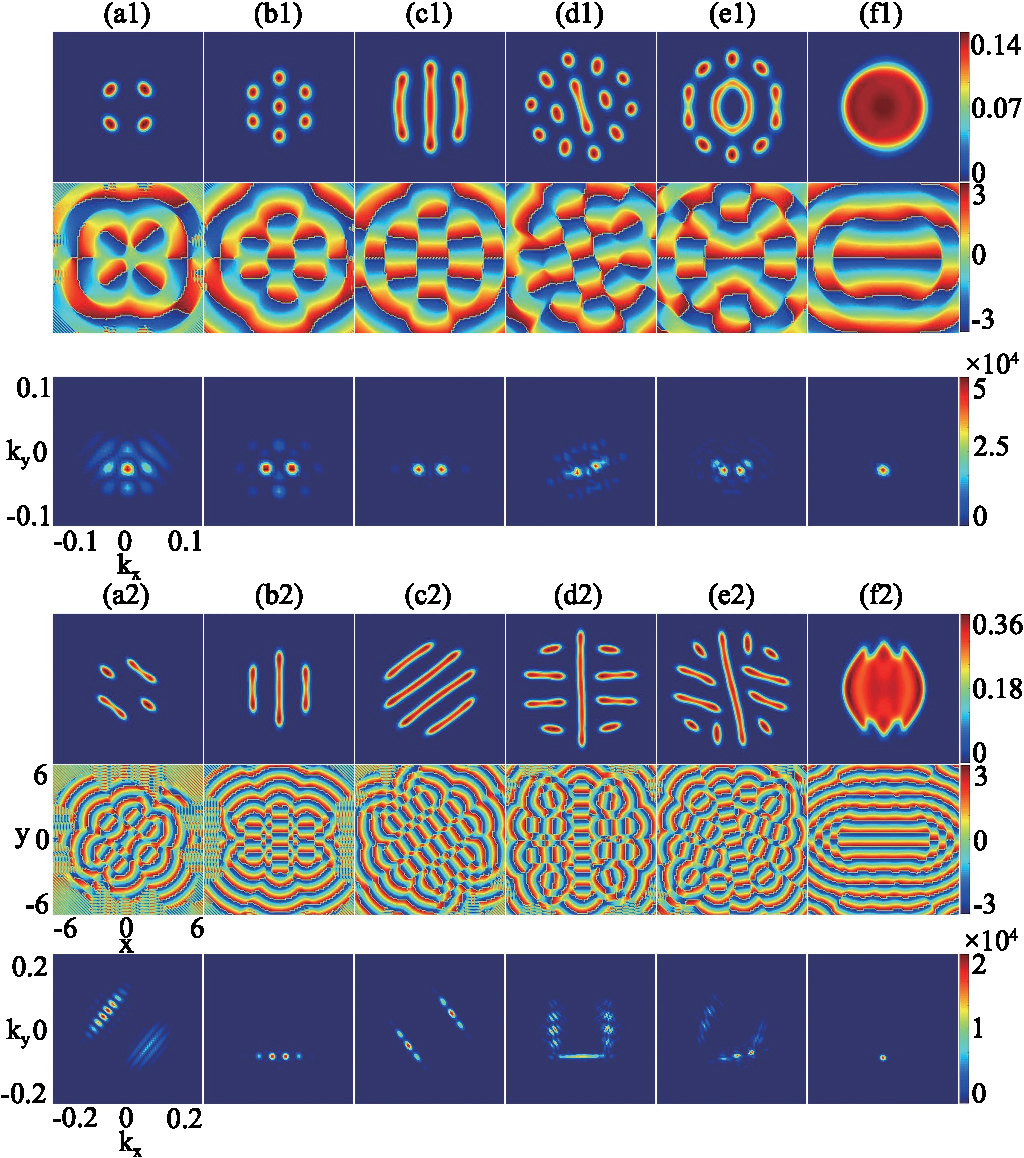}}
\caption{(Color online) Ground-state structures of dipolar spin-1/2 BECs with Rashba SOC and LHY corrections. The SOC strengths are (a1)-(f1) $\protect\kappa _{x}=\protect\kappa _{y}=2$ and (a2)-(f2) $\protect\kappa _{x}=\protect\kappa _{y}=6$, respectively. The other parameters in columns (a)-(f) (from left to right) are the same as those in Figs.2(a)-2(f). The first and fourth rows denote the density distributions of component 1, the second and fifth rows represent the phase distributions of component 1, and the third and sixth rows are the corresponding momentum distributions.}
\label{Figure6}
\end{figure}

From Fig. 6(c1), the QDs disappear and merge into three distinct elongated density stripes along the $y$ direction, with phase and momentum distributions similar to those in Fig. 6(b1). Hence the quantum phase in Fig. 6(c1) is a stripe phase with hidden vortex-antivortex cluster. However, for the D and E phases in Fig. 2, with the inclusion of weak SOC, the ground state of the system transforms into a droplet-stripe (annular stripe) coexisting phase with hidden vortex-antivortex cluster as shown in Figs. 6(d1) and 6(e1), where the momentum distributions become irregular. Finally, for the F phase (TF phase) in Fig. 2(f), the weak SOC does not cause substantial changes to the quantum phase structure of the system, and the ground state remains the TF phase (see Fig. 6(f1)). The small difference is that the momentum distribution in the $k$-space is concentrated at a point in the negative direction of the $k_{y}$ axis. Relevant studies show that the spin-orbit coupled system has an additional Zeeman term proportional to the velocity of the moving frame under the Galilean transformation \cite{Zhai,Sakaguchi}. This is a key feature of SOC that breaks Galilean invariance of the BECs. From the third row in Fig. 6, it can be seen that the momentum distribution of the system is mainly concentrated in the region of $k_{y}<0$. The physical reason is that the Galilean invariance in the spin-orbit coupled superfluid is broken, resulting in an uneven distribution of atomic velocity (momentum distribution) \cite{Sakaguchi,Gui}.

When the SOC strength is large, e.g., $\kappa _{x}=\kappa _{y}=6 $ (the bottom three rows in Fig. 6), the effect of SOC on the quantum phases of the system is more pronounced. As shown in Fig. 6(a2), the droplets are significantly stretched and adhered to each other, and the system tends to form a stripe phase with hidden vortex-antivortex cluster. The momentum distribution shows a special combination structure of a chain of scattered points and several stripes. In Figs. 6(b2)-6(e2), all the ground states become various stripe phases with hidden vortex-antivortex cluster, where the hidden vortex-antivortex pairs are mainly arranged along the directions of the stripes. In particular, the momentum distributions in Figs. 6(b2) and 6(c2) exhibit a discrete point chain and two parallel discrete point chains, respectively. Obviously, in view of the presence of these complicated topological defects, the stripe structures in Figs. 6(a2)-6(e2) are different from the conventional stripe phases in spin-orbit coupled BECs \cite{Spielman,Zhai,YZhang} due to the comprehensive competition among SOC, DDI, $s$-wave contact interaction, and LHY correction. In Fig. 6(f2), it is demonstrated that strong SOC tends to make the TF phase develop towards the stripe phase. The main physical reason is that SOC spontaneously breaks the space-spin rotation symmetry, making the system prone to exhibiting stripe distribution. In a word, for the present system, our results indicate that weak or moderate SOC is usually beneficial for promoting the formation of QDs, but strong SOC may destroy the QDs and lead to the generation of stripe phase.

\subsection{Spin textures}

\begin{figure*}[tbp]
\centerline{\includegraphics*[width=14cm]{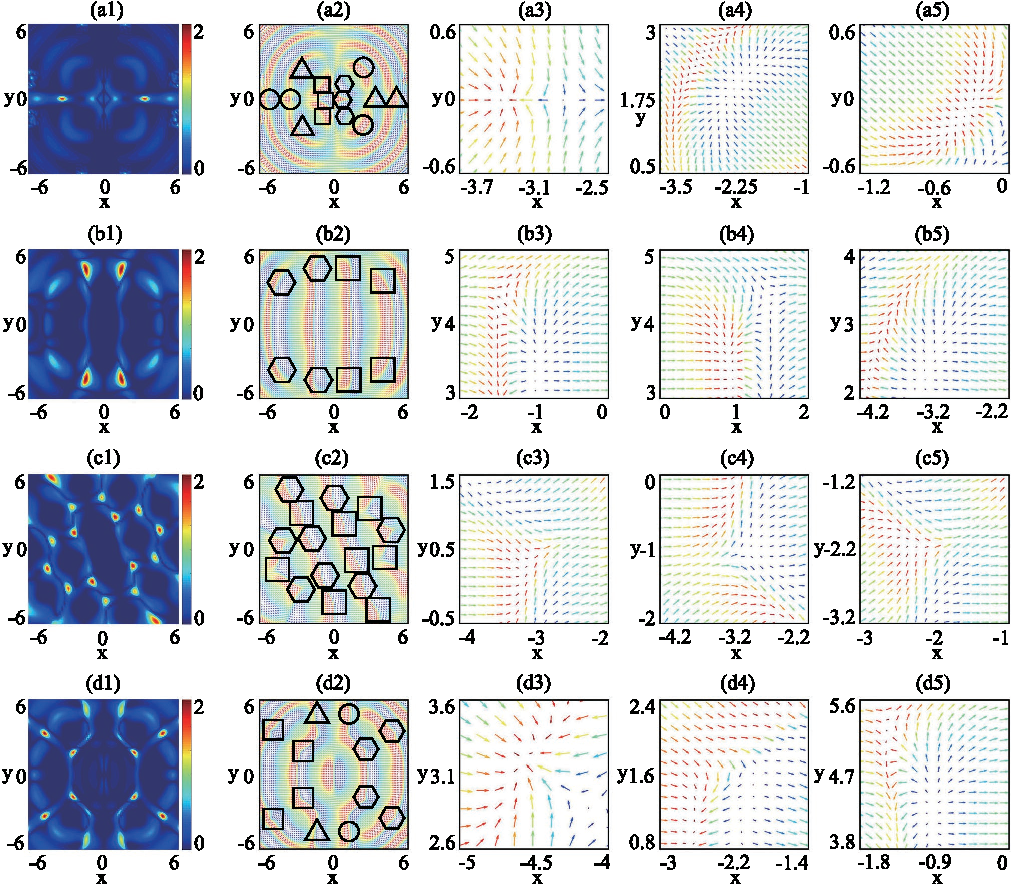}}
\caption{(Color online) Topological charge densities and spin textures of
dipolar spin-1/2 BECs with SOC and LHY corrections in a harmonic trap, where
the corresponding ground states for (a)-(d) are given in Figs. 6(a1), 6(c1), 6(d1)
and 6(e1), respectively. The arrows in the spin texture represent the
transverse spin vector ($S_{x}$, $S_{y}$) and the color of each arrow
indicates the magnitude of $S_{z}$. The first column (from left to right)
denotes topological charge density, the second column is spin texture,
and the right three columns represent the local amplifications of the spin
texture. The circle, triangle, square, and hexagon denote a skyrmion, an antiskyrmion, a
half-skyrmion (meron), and a half-antiskyrmion (antimeron), respectively. The unit length is $a_{0}$.}
\label{Figure7}
\end{figure*}

Now we analyze the topological charge densities and spin textures of dipolar BECs with SOC and LHY corrections in order to further elucidate the ground-state properties. In Fig. 7, we show the topological charge densities (column 1) and the
spin textures (column 2) of the system in the left two columns, and the
local amplifications of the spin textures are displayed in the right three
columns. The ground states for Figs. 7(a)-7(d) are given in Figs. 6(a1), 6(c1), 6(d1)
and 6(e1), respectively. For the sake of discussion, we use circle, triangle, square, and hexagon to denote a skyrmion \cite{Skyrme}, an antiskyrmion, a half-skyrmion (meron) \cite{Mermin}, and a half-antiskyrmion (antimeron), respectively.

Our computation results show that the local topological charges in Figs.
7(a3)-7(a5) approach $Q=1$, $Q=-1$ and $Q=0.5$, respectively, which indicates
that the local topological defects in Figs. 7(a3)-7(a5) are skyrmion,
antiskyrmion and meron. The local topological charges at the corresponding right-side symmetric positions are $Q=-1$, $Q=1$ and $Q=-0.5$, respectively, meaning that the corresponding topological defects at the right-side symmetric positions are antiskyrmion, skyrmion and antimeron, respectively. Meanwhile, the spin texture is symmetric concerning the $y=0$ axis and antisymmetric concerning the $x=0$ axis. Thus the topological structure in Fig. 7(a2) is an exotic composite skyrmion-antiskyrmion-meron-antimeron cluster composed of four skyrmion-antiskyrmion pairs and three meron-antimeron pairs. The spin defect in Fig. 7(b2) is a meron-antimeron string cluster consists of two meron strings and two antimeron strings, where the local enlargements of the spin texture are exhibited in Figs. 7(b3)-7(b5). Intriguingly, the spin density in Fig. 7(c1) shows an approximate diagonal distribution, forming four twisted chains. As can be seen from the spin texture in Fig. 7(c2), each twisted chain corresponds to an interlaced antimeron-meron-antimeron chain, where the local amplifications of the spin texture are displayed in Figs. 7(c3)-7(c5). Therefore the spin defects in Fig. 7(c2) constitute a rather complicated antimeron-meron-antimeron chain cluster. In particular, the topological charge density in Fig. 7(d1) displays even parity and excellent symmetry about the $x=0$ axis and the $y=0$ axis. The spin texture in Fig. 7(d2) forms a peculiar skyrmion-antiskyrmion-meron-antimeron necklace with a meron-antimeron necklace embedded inside and a spin Neel domain wall along the $x=0$ direction in the central region of the trap. The local enlargements of the spin texture are presented in Figs. 7(d3)-7(d5).

To the best of our knowledge, there are few reports on the study of spin
textures in the system with quantum fluctuations. These new skyrmion
excitations found in the present system are significantly different from
the skyrmion structures previously reported in other physical systems, for
instance, rotating two-component BECs with or without SOC (DDI) \cite%
{Lahaye,Zhai,WangH,XuXQ,Ramachandhran,Aftalion}. Furthermore, these novel
quantum phases and topological excitations (including vortex excitations and skyrmion
excitations) in present work allow to be tested and verified in the future experiments.

\section{Conclusions}

In summary, we have studied a rich variety of ground-state phases and topological excitations of quasi-2D
two-component dipolar BECs with LHY corrections and Rashba SOC in a harmonic
trap. In the absence of SOC, we present a ground-state phase diagram spanned by the DDI strength and the LHY correction strength. For given nonlinear interspecies
contact interactions, the system displays rich quantum phases, such as droplet lattice phase, annular phase, loop-island structure, stripe-droplet coexistence phase, toroidal stripe phase,
and TF phase. For the droplet lattice phase, the number of droplets shows an approximate linear dependence on the LHY correction coefficient. In addition, to describe the changing process between different quantum phases of the system, we calculate the average energy of per atom for different quantum phases. In the stable quantum phase region, as the LHY correction strength increases, the energy per atom increase overall and finally approaches a steady value. At the same time, the energy per atom gradually increases with the increasing of DDI strength for the TF phase and the toroidal stripe phase, while it non-monotonically decreases with the increasing of DDI strength for the stripe-droplet coexistence phase and the droplet lattice phase. For the rotating case, the variation of DDI strength can lead to a quantum phase transition between superfluid phase and supersolid phase, as well as a quantum phase transition between different superfluid phases, depending on the LHY correction strength.

In the presence of SOC, the QDs are obviously stretched along a certain direction, and hidden vortex-antivortex pairs (or clusters) are created in
each component. In this case, the system sustains unique and novel ground-state quantum phases including square stretched droplet lattice phase with hidden vortex-antivortex pairs, triangular droplet lattice phase with hidden vortex-antivortex cluster, stripe phase with hidden vortex-antivortex cluster, and droplet-stripe (or annular stripe) coexisting phase with hidden vortex-antivortex cluster. It is shown that weak or moderate SOC favors the formation of droplets, whereas for strong SOC the ground state of the system develops into a stripe phase with hidden vortex-antivortex cluster. Furthermore, the typical spin textures of the system are analyzed. We find that the system supports exotic topological excitations, such as composite skyrmion-antiskyrmion-meron-antimeron cluster, meron-antimeron string cluster, complicated antimeron-meron-antimeron chain cluster, and peculiar skyrmion-antiskyrmion-meron-antimeron necklace with a meron-antimeron necklace embedded inside and a spin Neel domain wall in the central region of the trap. These rich novel quantum phases and topological excitations in this system allow to be observed in the future cold atom experiments. The findings in the present work have enriched our new understanding for the exotic quantum states and topological excitations in cold atom physics and condensed matter physics.

\acknowledgments
This work was supported by the National Natural Science Foundation of China
(Grant Nos. 11475144 and 11047033), Hebei Natural Science Foundation (Grant
Nos. A2022203001, A2019203049 and A2015203037), Innovation Capability
Improvement Project of Hebei province (Grant No. 22567605H), and Research
Foundation of Yanshan University (Grant No. B846).


\begin{thebibliography}{99}

\bibitem{Griesmaier} Griesmaier A, Werner J, Hensler S, Stuhler J, Pfau T. Phys Rev Lett
2005;94:160401.

\bibitem{Lu} Lu M, Burdick NQ, Youn SH, Lev BL. Phys Rev Lett 2011;107:190401.

\bibitem{Aikawa} Aikawa K, Frisch A, Mark M, Baier S, Rietzler A, Grimm R, Ferlaino F. Phys
Rev Lett 2012;108:210401.

\bibitem{Lahaye} Lahaye T, Menotti C, Santos L, Lewenstein M, Pfau T. Rep Prog Phys
2009;72:126401.

\bibitem{Josserand} Josserand C, Pomeau Y, Rica S. Phys Rev Lett 2007;98:195301.

\bibitem{Ancilotto1} Roccuzzo SM, Ancilotto F. Phys Rev A 2019;99:041601(R).

\bibitem{Huang} Lee TD, Huang K, Yang CN. Phys Rev 1957;106:1135.

\bibitem{Toennies} Toennies JP, Vilesov AF. Angew Chem Int Ed 2004:43:2622.

\bibitem{Barranco} Barranco M, Guardiola R, Hern\'{a}ndez S, Mayol R, Navarro J, Pi M. J Low Temp Phys
2006;142:1.

\bibitem{Kadau} Kadau H, Schmitt M, Wenzel M, Wink C, Maier T, Ferrier-Barbut I, Pfau T. Nature (London)
2016;530:194.

\bibitem{Chomaz1} Chomaz L, Baier S, Mark MJ, W\"{a}chtler F, Santos L, Ferlaino F. Phys Rev X
2016;6:041039.

\bibitem{Ferrier-Barbut1} Ferrier-Barbut I, Kadau H, Schmitt M, Wenzel M, Pfau T. Phys Rev Lett
2016;116:215301.

\bibitem{Baillie} Baillie D, Blakie PB. Phys Rev Lett 2018;121:195301.

\bibitem{Rafa} O\l dziejewski R, G\'{o}recki W, Paw\l owski K, Rz\c{a}\.{z}ewski K. Phys Rev
Lett 2020;124:090401.

\bibitem{Abdel} Boudjem\^{a}a A. New J Phys 2019;21:093027.

\bibitem{Young} Young-S LE, Adhikari SK. Phys Rev A 2022;105:033311.

\bibitem{Modugno1} Tanzi L, Lucioni E, Fam\`{a} F, Catani J, Fioretti A, Gabbanini C, Bisset
RN, Santos L, Modugno G. Phys Rev Lett 2019;122:130405.

\bibitem{Langen} B\"{o}ttcher F, Schmidt J-N, Wenzel M, Hertkorn J, Guo M, Langen T, Pfau T. Phys
Rev X 2019;9:011051.

\bibitem{ChomazL} Chomaz L, Petter D, Ilzh\"{o}fer P, et al. Phys Rev X 2019;9:021012.

\bibitem{Modugno2} Tanzi L, Maloberti JG, Fioretti A, Gabbanini C, Modugno G. Science 2021;371:6534.

\bibitem{Roccuzzo} Roccuzzo SM, Gallem\'{\i} A, Recati A, Stringari S. Phys Rev Lett
2020;124:045702.

\bibitem{Stringari} \v{S}indik M, Recati A, Roccuzzo SM, Santos L, Stringari S. Phys Rev A
2022;106:L061303.

\bibitem{Petrov} Petrov DS. Phys Rev Lett 2015;115:155302.

\bibitem{Cabrera} Cabrera CR, Tanzi L, Sanz J, Naylor B, Thomas P, Cheiney P, Tarruell L. Science
2018;359:301.

\bibitem{Mishra} Mishra C, Santos L, Nath R. Phys Rev Lett 2020;124:073402.

\bibitem{Smith} Smith JC, Baillie D, Blakie PB. Phys Rev Lett 2021;126:025302.

\bibitem{Dong} Dong L, Liu D, Du Z, Shi K, Qi W. Phys Rev A 2022;105:033321.

\bibitem{Elhadj} Boudjem\^{a}a A, Elhadj KM. Chaos Solitons Fractals 2023;176:114133.

\bibitem{ZhangY} Zhang Y, Su X, Chen H, Hong Y, Li J, Wen L. Results Phys 2023;54:107067.

\bibitem{DuX} Du X, Fei Y, Chen X-L, Zhang Y. Phys Rev A 2023;108:033312.

\bibitem{LSantos} Bisset RN, Pe\~{n}a Ardila LA, Santos L. Phys Rev Lett 2021;126:025301.

\bibitem{Rakshit} Rakshit D, Karpiuk T, Zin P, Brewczyk M, Lewenstein M, Gajda M. New J Phys
2019;21:073027.

\bibitem{Ma} Ma Y, Peng C, Cui X. Phys Rev Lett 2021;127:043002.

\bibitem{Oktel} Yo\u{g}urt TA, Kele\c{s} A, Oktel M\"{O}. Phys Rev A 2022;105:043309.

\bibitem{LiZ} Li Z, Pan J-S, Liu WV. Phys Rev A 2019;100:053620.

\bibitem{Spielman} Lin Y-J, Jim\'{e}nez-Garc\'{\i}a K, Spielman IB. Nature (London) 2011;471:83.

\bibitem{Wu} Wu Z, Zhang L, Sun W, Xu X-T, Wang B-Z, Ji S-C, Deng Y, Chen S, Liu X-J, Pan J-W. Science
2016;354:83.

\bibitem{WangZY} Wang Z-Y, Cheng X-C, Wang B-Z, et al. Science 2021;372:271.

\bibitem{Rashba} Bychkov YA, Rashba EI. J Phys C: Solid State Phys 1984;17:6039.

\bibitem{Dresselhaus} Dresselhaus G. Phys Rev 1955;100:580.

\bibitem{Goldman} Goldman N, Juzeli\={u}nas G, \"{O}hberg P, Spielman IB. Rep Prog Phys
2014;77:126401.

\bibitem{Zhai} Zhai H. Rep Prog Phys 2015;78:026001.

\bibitem{Sakaguchi} Sakaguchi H, Li B, Malomed BA. Phys Rev E 2014;89:032920.

\bibitem{YZhang} Zhang Y, Mao L, Zhang C. Phys Rev Lett 2012;108:035302.

\bibitem{WangH} Wang H, Wen L, Yang H, Shi C, Li J. J Phys B: At Mol Opt Phys 2017;50:155301.

\bibitem{Ketterle} Li JR, Lee J, Huang W, Burchesky S, Shteynas B, Top FC, Jamison AO, Ketterle W. Nature
(London) 2017;543:91.

\bibitem{Sinha} Sinha S, Nath R, Santos L. Phys Rev Lett 2011;107:270401.

\bibitem{Hu} Hu J, Wang Q, Su X, Zhang Y, Wen L. Results Phys 2022;34:105238.

\bibitem{XuXQ} Xu X-Q, Han JH. Phys Rev Lett 2011;107:200401.

\bibitem{Achilleos} Achilleos V, Frantzeskakis DJ, Kevrekidis PG, Pelinovsky DE. Phys Rev Lett
2013;110:264101.

\bibitem{Xu} Xu Y, Zhang Y, Wu B. Phys Rev A 2013;87:013614.

\bibitem{ZhangYC} Zhang Y-C, Zhou Z-W, Malomed BA, Pu H. Phys Rev Lett 2015;115:253902.

\bibitem{Kartashov1} Kartashov YV, Torner L, Modugno M, Sherman EY, Malomed BA, Konotop VV. Phys Rev Res 2020;2:013036.

\bibitem{WangQ} Wang Q, Zhao W, Wen L. Results Phys 2021;25:104317.

\bibitem{Ramachandhran} Hu H, Ramachandhran B, Pu H, Liu X-J. Phys Rev Lett 2012;108:010402.

\bibitem{LiuCF} Liu C-F, Fan H, Zhang Y-C, Wang D-S, Liu W-M. Phys Rev A 2012;86:053616.

\bibitem{Xiaoling} Cui X-L. Phys Rev A 2018;98:023630.

\bibitem{Yongyao} Li Y, Luo Z, Liu Y, Chen Z, Huang C, Fu S, Tan H, Malomed BA. New J Phys
2017;19:113043.

\bibitem{XuSL} Xu S-L, Lei Y-B, Du J-T, Zhao Y, Hua R, Zeng J-H. Chaos Solitons Fractals
2022;164:112665.

\bibitem{PfauT} Chomaz L, Ferrier-Barbut I, Ferlaino F, Laburthe-Tolra B, Lev LB, Pfau T.
Rep Prog Phys 2023;86:026401.

\bibitem{Kartashov} Kartashov YV, Malomed BA, Tarruell L, Torner L. Phys Rev A 2018;98:013612.

\bibitem{Cidrim} Cidrim A, dos Santos FEA, Henn EAL, Macr\`{\i} T. Phys Rev A 2018;98:023618.

\bibitem{Sachdeva} Sachdeva R, Tengstrand MN, Reimann SM. Phys Rev A 2020;102:043304.

\bibitem{Dalfovo} Dalfovo F, Giorgini S, Pitaevskii LP, Stringari S. Rev Mod Phys 1999;71:463.

\bibitem{Ticknor} Ticknor C, Wilson RM, Bohn JL. Phys Rev Lett 2011;106:065301.

\bibitem{Fischer} Fischer UR. Phys Rev A 2006;73:031602(R).

\bibitem{Nath} Nath R, Pedri P, Santos L. Phys Rev Lett 2009;102:050401.

\bibitem{Aftalion} Aftalion A, Mason P. Phys Rev A 2013;88:023610.

\bibitem{Kasamatsu2} Kasamatsu K, Tsubota M, Ueda M. Phys Rev A 2005;71:043611.

\bibitem{LiJ} Wen L, Li J. Phys Rev A 2014;90:053621.

\bibitem{Mardonov} Mardonov S, Sherman EY, Muga JG, Wang H-W, Ban Y, Chen X. Phys Rev A 2015;91:043604.

\bibitem{Shamriz} Shamriz E, Chen Z, Malomed BA. Phys Rev A 2020;101:063628.

\bibitem{Bisset} Bisset RN, Wilson RM, Baillie D, Blakie PB. Phys Rev A 2016;94:033619.

\bibitem{Sean} O'Connell SMO, Tanyag RMP, Verma D, et al. Phys Rev Lett 2020;124:215301.

\bibitem{WenL1} Wen L, Xiong H, Wu B. Phys Rev A 2010;82:053627.

\bibitem{WenL2} Wen LH, Luo XB. Laser Phys Lett 2012;9:618.

\bibitem{Rica} Pomeau Y, Rica S. Phys Rev Lett 1994;72:2426.

\bibitem{Pohl} Macr\`{\i}T, Maucher F, Cinti F, Pohl T. Phys Rev A 2013;87:061602(R).

\bibitem{Shlyapnikov} Lu Z-K, Li Y, Petrov DS, Shlyapnikov GV. Phys Rev Lett 2015;115:075303.

\bibitem{Gui} Gui Z, Zhang Z, Su J, Lyu H, Zhang Y. Phys Rev A 2023;108:043311.

\bibitem{Fetter} Fetter AL. Rev Mod Phys 2009;81:647.

\bibitem{Ueda} Kasamatsu K, Tsubota M, Ueda M. Phys Rev A 2003;67:033610.

\bibitem{Skyrme} Skyrme THR. Nucl Phys 1962;31:556.

\bibitem{Mermin} Mermin ND, Ho T-L. Phys Rev Lett 1976;36:594.

\end{thebibliography}
\end{document}